\documentclass [lettersize,journal]{IEEEtran}
\usepackage{amsmath,amssymb,amscd}
\usepackage{algorithm}
\usepackage{array}
\usepackage[caption=false,font=normalsize,labelfont=sf,textfont=sf]{subfig}
\usepackage{textcomp}
\usepackage{stfloats}
\usepackage{url}
\usepackage{verbatim}
\usepackage{graphicx}
\usepackage{cite}
\usepackage{bbold}
\usepackage{algpseudocode}
\usepackage{multirow}
\usepackage{amsmath,systeme}
\usepackage{mathtools}
\usepackage{xcolor}
\usepackage{tikz}
\usepackage{orcidlink}
\usepackage{lipsum}
\usepackage{cuted}
\usepackage[belowskip=-2pt,aboveskip=5pt,font=footnotesize, justification=raggedright, singlelinecheck=false]{caption}
\usepackage{caption}
\usepackage[labelsep=period]{caption}
\usepackage{comment}

\newcommand{\RNum}[1]{\uppercase\expandafter{\romannumeral #1\relax}}
\renewcommand{\arraystretch}{0.6}
\newcommand{\rightvsarrow}[1][3pt]{\mathrel{%
   \hbox{\rule[\dimexpr\fontdimen22\textfont2-.2pt\relax]{#1}{.4pt}}%
   \mkern-5mu\hbox{\usefont{U}{lasy}{m}{n}\symbol{41}}}}

\newtheorem{lemma}{Lemma}
\begin{document}

\title{Joint Mechanical and Electrical Adjustment of IRS-aided LEO Satellite MIMO Communications}

\author{Doyoung Kim\orcidlink{0009-0008-1098-9111},~\IEEEmembership{Student Member,~IEEE}, and Seongah Jeong\orcidlink{0000-0002-9737-0432},~\IEEEmembership{Senior Member,~IEEE}
        % <-this % stops a space

\vspace{-0.5cm}

\thanks{This work was supported by the National Research Foundation of Korea (NRF) grant funded by the Korea government (MSIT) (No. 2023R1A2C2005507). 
(\textit{Corresponding author}: Seongah Jeong.)}% <-this % stops a space
\thanks{Doyoung Kim and Seongah Jeong are with the School of Electronic and Electrical Engineering, Kyungpook National University, Daegu 41566, South Korea (Email: singha5036@knu.ac.kr and seongah@knu.ac.kr).}}
\markboth{ }%
{Shell \MakeLowercase{\textit{et al.}}: A Sample Article Using IEEEtran.cls for IEEE Journals}


\maketitle
\begin{abstract}
In this correspondence, we propose a joint mechanical and electrical adjustment of intelligent reflecting surface (IRS) for the performance improvements of low-earth orbit (LEO) satellite multiple-input multiple-output (MIMO) communications.
In particular, we construct a three-dimensional (3D) MIMO channel model for the mechanically-tilted IRS in general deployment, and consider two types of scenarios with and without the direct path of LEO-ground user link due to the orbital flight. With the aim of maximizing the end-to-end performance, we jointly optimize tilting angle and phase shift of IRS along with the transceiver beamforming, whose performance superiority is verified via simulations with the Orbcomm LEO satellite using a real orbit data.
\end{abstract}
\vspace{-0.1cm}

\vspace{-0.4cm}

\section{Introduction}\label{sec:intro}
\IEEEPARstart{W}{ith} the rapidly impending sixth-generation (6G) wireless era, low-earth-orbit (LEO) satellite communications have emerged as a prospective solution to alleviate global coverage disparities, supported by global satellite companies such as Starlink, OneWeb, and Orbcomm \cite{Kassas23PLANS}.
In accordance with the development trends for coverage enhancements, intelligent reflecting surface (IRS) has been explored to enable the seamless LEO communications by establishing the virtual links beyond the visible time window and to compensate the path loss inherent in long-distance transmission of LEO satellites \cite{Matthiesen21WCL, Zheng22JSAC, Tian22arXiv, Zheng22GCWkshps, Cao23PIMRC}. The authors in \cite{Matthiesen21WCL} leverage the satellite's predictable trajectory to design the optimal phase shift at IRS for maximizing the signal-to-noise ratio (SNR). In \cite{Zheng22JSAC}, the three-dimensional (3D) multiple-input multiple-output (MIMO) channel model is investigated with the assistance of IRS installed on both LEO satellite and ground user (GU).

Recently, the potential of mechanically-tilted IRS has been discussed as a means to enhance conventional terrestrial networks \cite{Cheng22TC, Tang21WC, Cheng22TWC, Hadzi21WCL, Wang22ICCT}, as well as to improve the coverage of LEO satellite communication in urban scenarios \cite{Tian22arXiv, Zheng22GCWkshps, Cao23PIMRC}. In \cite{Zheng22GCWkshps}, the joint optimization of passive beamforming and orientation of the tilted IRS is developed to maximize the average throughput of GUs. The authors in \cite{Cao23PIMRC} consider the near-field channel model between the GU and the IRS and optimize the passive beamforming and orientation of the tilted IRS to maximize the received power at GU. The benefits of mechanically-tilted IRS can be further enhanced by the aligned MIMO at transceiver, which necessitates the development of an intricate 3D channel modeling. Previous works on tilted IRS design \cite{Cheng22TC, Tang21WC, Cheng22TWC, Hadzi21WCL, Wang22ICCT, Tian22arXiv, Zheng22GCWkshps, Cao23PIMRC} mostly assume a single antenna at the transceivers, with the center elements of both the transceivers and the tilted IRS located on the same plane.

\begin{figure}[t!]
\centering
\includegraphics[width=6.0cm]{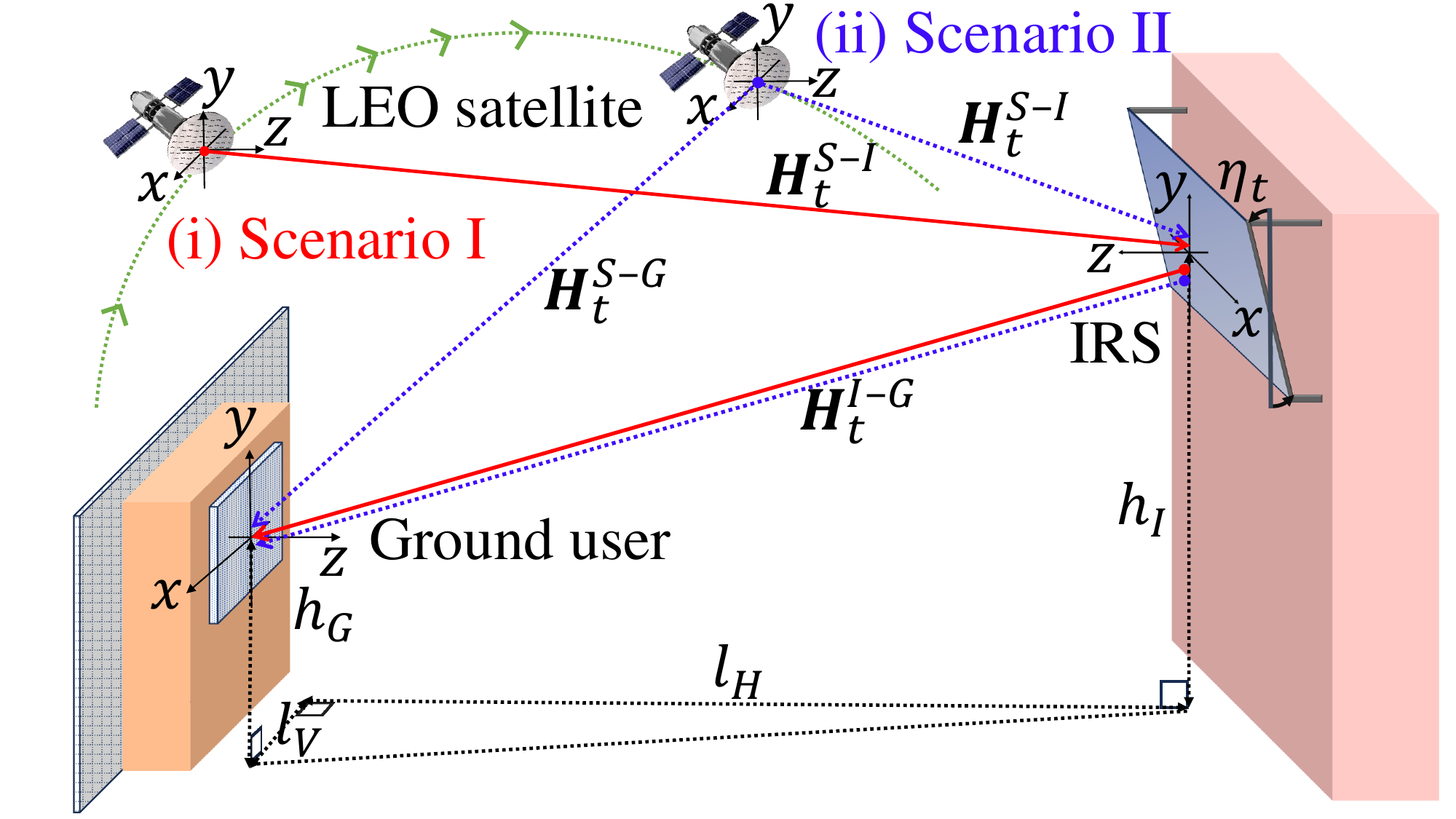}
\caption{System model of IRS-aided LEO satellite communications.}
\label{fig:sys_1}
    \vspace{-0.4cm}
\end{figure}

In this correspondence, we propose the joint mechanical and electrical design of mechanically-tilted IRS for the LEO satellite MIMO communications. To this end, a 3D MIMO channel modeling is proposed for the general deployment of the tilted IRS in real applications. The two types of possible LEO communications are studied with and without direct link between LEO satellite and GU, owing to the predetermined orbital flight of the LEO satellite and the surroundings of GU.
To improve the end-to-end performance in terms of signal-to-noise ratio (SNR), we jointly optimize both tilting angle and phase shift matrix of IRS along with the transceiver beamforming for each scenario. For practical verification, the performance of proposed joint design is analyzed with the real orbit data of Orbcomm LEO satellite compared to the existing solutions.
To the best of our knowledge, this is the first attempt to develop the joint mechanical and electrical design for the general IRS deployment in LEO satellite MIMO systems.

    \vspace{-0.2cm}
\section{System Model}\label{sec:sys}

\begin{figure}[t] 
\centering
{
    \includegraphics[width=0.23\textwidth] {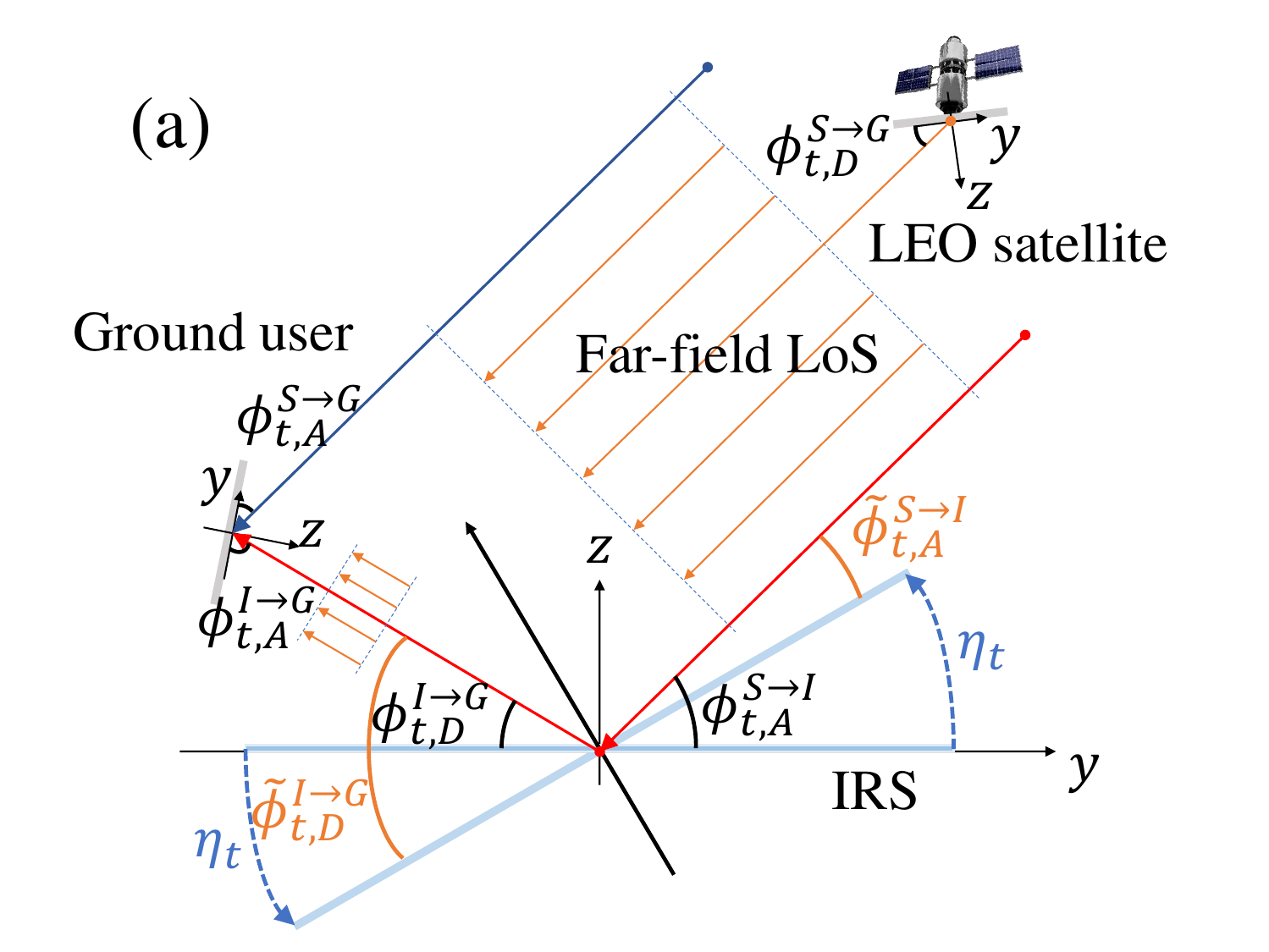}%
    \label{fig:sys2}
}
{
    \includegraphics[width=0.23\textwidth] {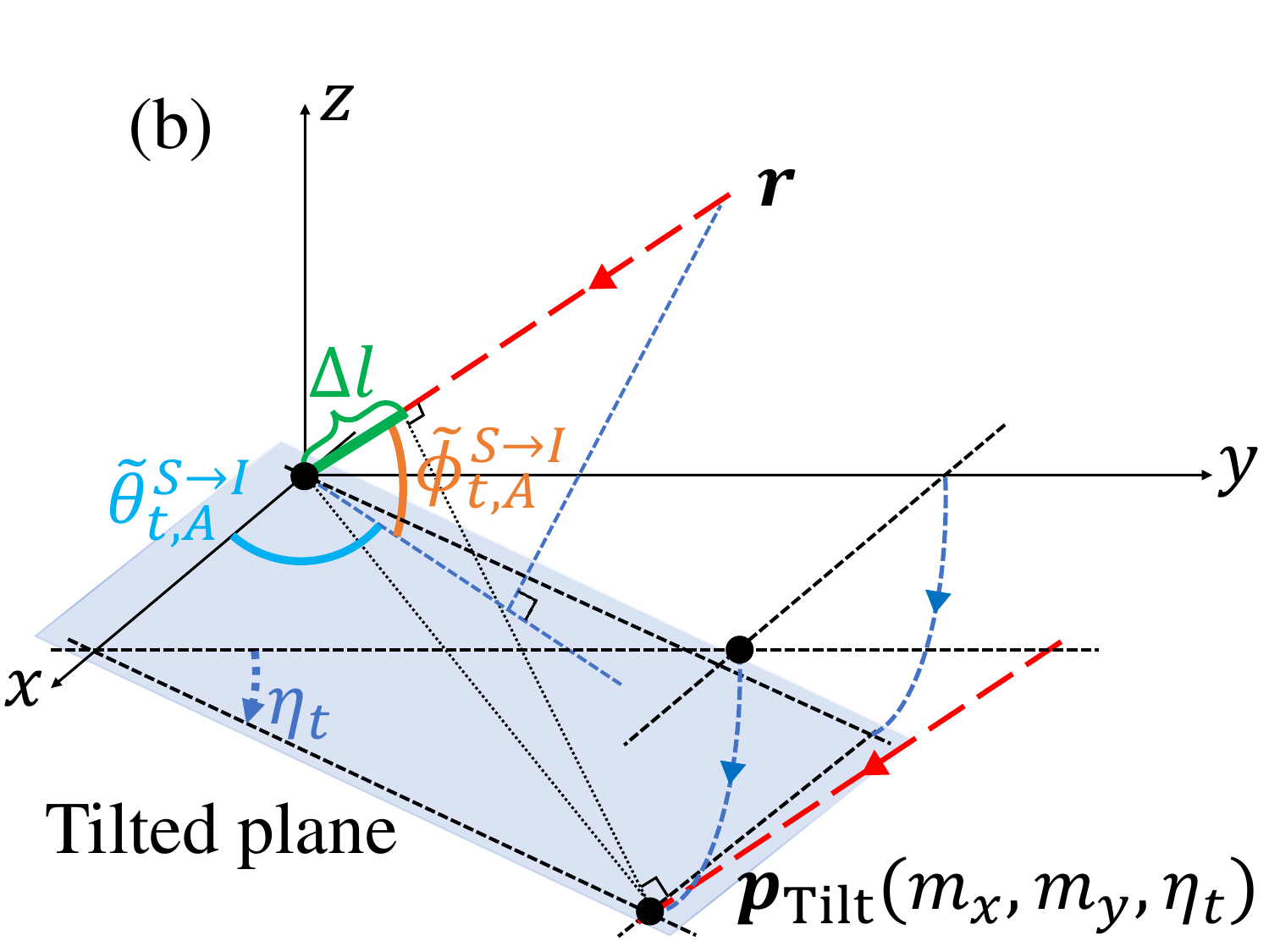}%
    \label{fig:sys3}
}
\caption{Illustration of angular relationship with tilting angle of $\eta_t$. (a) 2D illustration of tilted IRS (b) 3D illustration of tilted IRS}
\label{fig:sys23}
    \vspace{-0.5cm}
\end{figure}

We consider an IRS-aided LEO satellite communication system for downlink as illustrated in Fig. \ref{fig:sys_1}, consisting of a GU, a tilted IRS and a LEO satellite, all of which are supposed to be equipped with the uniform planar arrays (UPAs) along $x$-axis (horizontal) and $y$-axis (vertical) of their respective planes. 
We assume that the heights of IRS and GU's antenna are $h_I$ \hspace{-0.1cm}(m) and $h_G$ \hspace{-0.1cm}(m), respectively, and the horizontal distance $l_H$ \hspace{-0.1cm}(m) and the vertical distance $l_V$ \hspace{-0.1cm}(m) are considered between the centers of IRS and GU as in Fig. \ref{fig:sys_1}.
For simplification, we use the indices of $G$, $S$ and $I$ for parameters related to GU, LEO satellite and IRS, respectively.
The UPAs of GU and LEO satellite consist of $N_{i}\hspace{-0.05cm}\triangleq\hspace{-0.05cm} N_{x, i} \hspace{-0.05cm}\times  \hspace{-0.05cm}N_{y, i}$ active antenna elements along the $x$-axis and the $y$-axis with the spacing of $d_{x,i}$ and $d_{y,i}$, respectively, where $i \hspace{-0.05cm}\in \hspace{-0.05cm} \{G, S\}$.
Similarly, the IRS is composed of $M \triangleq M_{x} \times M_{y}$ passive reflecting elements with its spacings $d_{x,I}$ and $d_{y,I}$. The phase shift matrix of IRS can be represented as $\pmb{\Theta}_t\triangleq \text{diag}\left(\theta_t^1, \theta_t^2, \cdots \theta_t^M \right)$, where the reflection amplitudes of all reflecting elements are set to be one without loss of generality, i.e., $\lvert \theta_t^m \rvert = 1$, for $m \hspace{-0.05cm}\in \hspace{-0.05cm}\{1,\dots,M\}$ \cite{Zheng22GCWkshps}.
Due to the extremely long inter-node distance, the channels can be characterized by the far-field line-of-sight (LoS) model with parallel wavefronts according to the definition of Rayleigh distance \cite{Zheng22JSAC, Tian22arXiv, Zheng22GCWkshps}.
For channel modeling, we use the subscripts of $D$ and $A$ for the angle of departure (AoD) and the angle of arrival (AoA), respectively.
In the downlink channel, the pairs of $\pmb{\psi}_{t,D}^{S \rightvsarrow \hspace{-0.05cm} X}\hspace{-0.05cm}=\hspace{-0.05cm}(\theta_{t,D}^{S \rightvsarrow \hspace{-0.05cm} X}, \phi_{t,D}^{S \rightvsarrow \hspace{-0.05cm} X})$ and $\pmb{\psi}_{t,A}^{S \rightvsarrow \hspace{-0.05cm} X}\hspace{-0.05cm}=\hspace{-0.05cm}(\theta_{t,A}^{S \rightvsarrow \hspace{-0.05cm} X}, \phi_{t,A}^{S \rightvsarrow \hspace{-0.05cm} X})$ are defined as the AoD at the LEO satellite and the AoA at the GU and IRS, for $X\in\{G,I\}$, based on the center of plane without tilting at IRS.
Similarly, we define $\pmb{\psi}_{t,D}^{I \rightvsarrow \hspace{-0.02cm} G}\hspace{-0.05cm}=\hspace{-0.05cm}(\theta_{t,D}^{I \rightvsarrow \hspace{-0.02cm} G}, \phi_{t,D}^{I \rightvsarrow \hspace{-0.02cm} G})$ and $\pmb{\psi}_{t,A}^{I \rightvsarrow \hspace{-0.02cm} G}\hspace{-0.05cm}=\hspace{-0.05cm}(\theta_{t,A}^{I \rightvsarrow \hspace{-0.02cm} G}, \phi_{t,A}^{I \rightvsarrow \hspace{-0.02cm} G})$ for the  AoD at IRS and the AoA at GU, respectively,  (see, Fig. \ref{fig:sys23}). 
Considering the relative positioning of GU and LEO satellite with respect to the IRS plane, the azimuth angles can be specified as $-\pi\hspace{-0.1cm} <\hspace{-0.1cm}\theta_{t,D}^{I \rightvsarrow \hspace{-0.02cm} G}\hspace{-0.1cm}<\hspace{-0.1cm}0$ and
 $0\hspace{-0.1cm}<\hspace{-0.1cm}\theta_{t,A}^{S \rightvsarrow \hspace{-0.02cm} I}\hspace{-0.1cm}<\hspace{-0.1cm}\pi$.
According to the time-variant positions of LEO satellite due to its orbital flight, two types of scenarios can be possible such as (i) Scenario I without direct link between LEO satellite and GU, where the proposed mechanical and electrical IRS design is indispensable and (ii) Scenario II with direct link, where LEO satellite communications can be enhanced thanks to the IRS.
For all nodes, the array response of the UPA can be represented by the Kronecker product of two uniform linear array (ULA) response vectors along the $x$-axis and the $y$-axis of each nodes. 
To this end, we start by introducing a one-dimensional (1D) steering vector for the ULA as 
\vspace{-0.1cm}
\begin{eqnarray}\label{eq:steering}
\pmb{v}(\delta, N) \triangleq \left[e^{-jk_w\lfloor-(N/2)+1\rfloor \delta}, \cdots ,1, \cdots , e^{jk_w\lfloor N/2 \rfloor \delta} \right]^T,
\end{eqnarray}
where $\delta$ indicates the path difference of the incoming signals between two adjacent elements, $N$ represents the number of elements, and $k_w=2\pi/\lambda$ is wavenumber defined by the signal wavelength $\lambda$ \cite{Matthiesen21WCL}. 
Accordingly, the array responses of LEO satellite and GU in LEO satellite-IRS and LEO satellite-GU links can be defined as
\begin{eqnarray}
&&\hspace{-1.2cm}\pmb{a}_{t,D}^{S \rightvsarrow \hspace{-0.05cm} X}(\pmb{\psi}_{t,D}^{S \rightvsarrow \hspace{-0.05cm} X}) = \pmb{v}(d_{x,S} \cos \phi_{t,D}^{S \rightvsarrow \hspace{-0.05cm} X}\cos \theta_{t,D}^{S \rightvsarrow \hspace{-0.05cm} X},  N_{x,S}) \nonumber \\
&&\hspace{+0.5cm} \otimes \hspace{0.1cm} \pmb{v}(d_{y,S} \cos \phi_{t,D}^{S \rightvsarrow \hspace{-0.05cm} X} \sin \theta_{t,D}^{S \rightvsarrow \hspace{-0.05cm} X},  N_{y,S} )\in \mathbb{C}^{N_{S}\times1}, \label{eq:arrayresp_S}\\
&&\hspace{-1.2cm}\pmb{a}_{t,A}^{S \rightvsarrow \hspace{-0.02cm} G}(\pmb{\psi}_{t,A}^{S \rightvsarrow \hspace{-0.02cm} G}) = \pmb{v}(d_{x,G} \cos \phi_{t,A}^{S \rightvsarrow \hspace{-0.02cm} G}\cos \theta_{t,A}^{S \rightvsarrow \hspace{-0.02cm} G},  N_{x,G}) \nonumber \\
&&\hspace{+0.5cm} \otimes \hspace{0.1cm} \pmb{v}(d_{y,G} \cos \phi_{t,A}^{S \rightvsarrow \hspace{-0.02cm} G} \sin \theta_{t,A}^{S \rightvsarrow \hspace{-0.02cm} G},  N_{y,G} )\in \mathbb{C}^{N_{G}\times1}, \label{eq:arrayresp_X}
\end{eqnarray} 
for $X\in\{G,I\}$, respectively. Also, the array response of GU in the IRS-GU link can be obtained as
\begin{eqnarray}
&&\hspace{-1.2cm}\pmb{a}_{t,A}^{I \rightvsarrow \hspace{-0.02cm} G}(\pmb{\psi}_{t,A}^{I \rightvsarrow \hspace{-0.02cm} G}) = \pmb{v}(d_{x,G} \cos \phi_{t,A}^{I \rightvsarrow \hspace{-0.02cm} G}\cos \theta_{t,A}^{I \rightvsarrow \hspace{-0.02cm} G},  N_{x,G})\nonumber \\
&&\hspace{+0.5cm} \otimes \hspace{0.1cm} \pmb{v}(d_{y,S} \cos \phi_{t,A}^{I \rightvsarrow \hspace{-0.02cm} G} \sin \theta_{t,A}^{I \rightvsarrow \hspace{-0.02cm} G},  N_{y,G} )\in \mathbb{C}^{N_{G}\times1}. \label{eq:arrayresp_G}
\end{eqnarray}
For the 3D array response of the mechanically-tilted IRS, we denote the coordinate of each passive elements as $\pmb{p}_{o}(m_x,m_y)$$=$$[d_{x,I}m_x, d_{y,I}m_y,0]^T$ with $m_x \in \{\lfloor -{M_x}/{2} +1 \rfloor ,\cdots, 1, \cdots, \lfloor {M_x}/{2}\rfloor \}$ and $m_y \in \{\lfloor -{M_y}/{2} +1 \rfloor , \cdots \, 1, \cdots, \lfloor {M_y}/{2}\rfloor \}$ along its $x$-axis and the $y$-axis without tilting. 
When the IRS is tilted with respect to the $y$-axis by $\eta_{t}$ (degrees), the path difference $\Delta l(\pmb{\psi}_{t,A}^{S \rightvsarrow \hspace{-0.02cm}I}, \eta_{t})$ of each passive element compared to the center element can be obtained as
\begin{eqnarray}\label{eq:pathdiff}
&&\hspace{-0.7cm}\Delta l (\pmb{\psi}_{t,A}^{S \rightvsarrow \hspace{-0.02cm}I}, \eta_{t}) = {\pmb{r}}(\pmb{\psi}_{t,A}^{S \rightvsarrow \hspace{-0.02cm}I})\cdot {\pmb{p}}_{\text{Tilt}}(m_x,m_y,\eta_{t}) \nonumber \\
&&\hspace{-0.7cm}= m_x\cos\phi_{t,A}^{S \rightvsarrow \hspace{-0.02cm} I}\cos\theta_{t,A}^{S \rightvsarrow \hspace{-0.02cm} I}d_{x,I}+m_y\cos\phi_{t,A}^{S \rightvsarrow \hspace{-0.02cm} I}\sin\theta_{t,A}^{S \rightvsarrow \hspace{-0.02cm} I}\cos \eta_{t} d_{y,I}\nonumber \\
&&\hspace{-0.4cm}+m_y\sin\phi_{t,A}^{S \rightvsarrow \hspace{-0.02cm} I}\sin\eta_{t}d_{y,I},
\end{eqnarray} where $\pmb{r}(\pmb{\psi}_{t,A}^{S \rightvsarrow \hspace{-0.02cm}I}) \hspace{-0.10cm}=\hspace{-0.10cm} (\hspace{-0.02cm}\cos \phi_{t,A}^{S \rightvsarrow \hspace{-0.02cm} I} \cos \theta_{t,A}^{S \rightvsarrow \hspace{-0.02cm} I}, \cos \phi_{t,A}^{S \rightvsarrow \hspace{-0.02cm} I}\sin\theta_{t,A}^{S \rightvsarrow \hspace{-0.02cm} I},\sin \phi_{t,A}^{S \rightvsarrow \hspace{-0.02cm} I}\hspace{-0.02cm})$ is the unit vector of the incident signal, and $\pmb{p}_{\text{Tilt}}(m_x,m_y,\eta_{t})$ can be induced by the rotation matrix $\pmb{R}(\eta_{t})$ and $\pmb{p}_{o}(m_x, m_y)$ as
\vspace{-0.2cm}
\begin{eqnarray}\label{eq:rotation}
&&\hspace{-1.2cm}
{\pmb{p}}_{\text{Tilt}}(m_x,m_y,\eta_{t})=
\pmb{R}(\eta_{t}) \pmb{p}_o(m_x, m_y) \nonumber\\
&&\hspace{+1.3cm}=\begin{bmatrix}
1 & 0 &  0 \\ 0 & \cos\eta_{t} &  -\sin\eta_{t} \\ 0 & \sin\eta_{t} & \ \ \cos\eta_{t} 
\end{bmatrix}
\begin{bmatrix}
m_x d_{x,I}\\ m_y d_{y,I} \\ 0
\end{bmatrix}.
\end{eqnarray}
By applying (\ref{eq:steering}) to (\ref{eq:pathdiff}), the array response vectors of the tilted IRS by $\eta_t$ can be expressed as
\begin{eqnarray}\label{eq:arrayresp_I_1}
&&\hspace{-1.2cm}\tilde{\pmb{a}}_{t,A}^{S \rightvsarrow \hspace{-0.02cm} I}(\pmb{\psi}_{t,A}^{S \rightvsarrow \hspace{-0.02cm}I}, \eta_{t}) = \pmb{v}(d_{x,I} \cos \phi_{t,A}^{S \rightvsarrow \hspace{-0.02cm} I} \cos \theta_{t,A}^{S \rightvsarrow \hspace{-0.02cm} I},  M_{x}) \otimes \nonumber\\
&&\hspace{-0.8cm}\pmb{v}(d_{y,I}( \cos \phi_{t,A}^{S \rightvsarrow \hspace{-0.02cm} I} \sin\theta_{t,A}^{S \rightvsarrow \hspace{-0.02cm} I} \cos \eta_{t} + \sin\phi_{t,A}^{S \rightvsarrow \hspace{-0.02cm} I} \sin \eta_{t}),  M_{y}), \\
&&\hspace{-1.2cm}\tilde{\pmb{a}}_{t,D}^{I \rightvsarrow \hspace{-0.02cm} G}(\pmb{\psi}_{t,D}^{I \rightvsarrow \hspace{-0.02cm}G}
, \eta_{t}) = \pmb{v}(d_{x,I} \cos \phi_{t,D}^{I \rightvsarrow \hspace{-0.02cm} G} \cos \theta_{t,D}^{I \rightvsarrow \hspace{-0.02cm} G},  M_{x}) \otimes \nonumber\\
&&\hspace{-0.8cm}\pmb{v}(d_{y,I}( \cos \phi_{t,D}^{I \rightvsarrow \hspace{-0.02cm} G} \sin \theta_{t,D}^{I \rightvsarrow \hspace{-0.02cm} G} \cos \eta_{t} + \sin \phi_{t,D}^{I \rightvsarrow \hspace{-0.02cm} G} \sin \eta_{t}),  M_{y}).\label{eq:arrayresp_I_2}
\end{eqnarray}
\hspace{-0.2cm}
With respect to the tilted plane of IRS, the modified elevation angles $\tilde{\phi}_{t,A}^{S \rightvsarrow \hspace{-0.02cm} I}(\pmb{\psi}_{t,A}^{S \rightvsarrow \hspace{-0.02cm}I}, \eta_{t})$ and $\tilde{\phi}_{t,D}^{I \rightvsarrow \hspace{-0.02cm} G}(\pmb{\psi}_{t,D}^{I \rightvsarrow \hspace{-0.02cm}G}, \eta_{t})$ can be calculated by the fact that tilting the IRS by $\eta_{t}$ can be equivalently interpreted as rotating the LEO satellite and the GU in the reverse direction by $-\eta_{t}$ in the original plane of IRS without tilting. Accordingly, the resultant unit vector of incident signal at IRS can be represented as 
\begin{eqnarray}\label{app:rotation}
&&\hspace{-1.2cm}\tilde{\pmb{r}}(\pmb{\psi}_{t,A}^{S \rightvsarrow \hspace{-0.02cm}I}, \eta_{t})\hspace{-0.05cm}=\hspace{-0.05cm} \pmb{R}(-\eta_{t}) \hspace{+0.05cm} \pmb{r}(\pmb{\psi}_{t,A}^{S \rightvsarrow \hspace{-0.02cm}I})\nonumber \\ 
&&\hspace{+0.45cm}=
\begin{bmatrix}
\cos \phi_{t,A}^{S \rightvsarrow \hspace{-0.02cm} I} \cos \theta_{t,A}^{S \rightvsarrow \hspace{-0.02cm} I}\\ 
\cos \phi_{t,A}^{S \rightvsarrow \hspace{-0.02cm} I} \sin \theta_{t,A}^{S \rightvsarrow \hspace{-0.02cm} I}\cos\eta_{t}+ \sin \eta_{t}\sin \phi_{t,A}^{S \rightvsarrow \hspace{-0.02cm} I}\\ 
\sin \phi_{t,A}^{S \rightvsarrow \hspace{-0.02cm} I} \cos \eta_{t} \hspace{-0.05cm}-\hspace{-0.05cm} \cos \phi_{t,A}^{S \rightvsarrow \hspace{-0.02cm} I} \sin \theta_{t,A}^{S \rightvsarrow \hspace{-0.02cm} I} \sin \eta_{t}
\end{bmatrix}^T \hspace{-0.3cm}.
\vspace{-0.6cm}
\end{eqnarray}
Based on $\tilde{\pmb{r}}(\pmb{\psi}_{t,A}^{S \rightvsarrow \hspace{-0.02cm}I}, \eta_{t})$ in (\ref{app:rotation}), the modified elevation angle 
$\tilde{\phi}_{t,A}^{S \rightvsarrow \hspace{-0.02cm} I} (\pmb{\psi}_{t,A}^{S \rightvsarrow \hspace{-0.02cm}I}, \eta_{t})$ of LEO satellite-IRS link can be expressed as
\begin{eqnarray}\label{app:Omega_S}
&&\hspace{-1.0cm}\tilde{\phi}_{t,A}^{S \rightvsarrow \hspace{-0.02cm} I}\hspace{-0.05cm}(\pmb{\psi}_{t,A}^{S \rightvsarrow \hspace{-0.02cm}I}, \eta_{t}) \hspace{-0.1cm}=\hspace{-0.1cm} \tan^{-1}\hspace{-0.1cm}\left(\hspace{-0.1cm} \frac{\sin(\alpha_{t,A}^{S \rightvsarrow \hspace{-0.02cm} I}-\eta_t)}{\sqrt{B_{t,A}^{S \rightvsarrow \hspace{-0.02cm} I}+\cos^2 (\alpha_{t,A}^{S \rightvsarrow \hspace{-0.02cm} I}-\eta_t)}}\hspace{-0.1cm}\right)\hspace{-0.1cm},
\end{eqnarray}
where $\alpha_{t,A}^{S \rightvsarrow \hspace{-0.02cm} I}\hspace{-0.12cm}=\hspace{-0.12cm}\tan^{-1}( {\tan \phi_{t,A}^{S \rightvsarrow \hspace{-0.02cm} I}}/{\sin \theta_{t,A}^{S \rightvsarrow \hspace{-0.02cm} I}} )$ and $B_{t,A}^{S \rightvsarrow \hspace{-0.02cm} I}\hspace{-0.12cm}=\hspace{-0.1cm}(\cos \phi_{t,A}^{S \rightvsarrow \hspace{-0.02cm} I}$ $\cos \theta_{t,A}^{S \rightvsarrow \hspace{-0.02cm} I})^2 /({(\cos \phi_{t,A}^{S \rightvsarrow \hspace{-0.02cm} I} \sin \theta_{t,A}^{S \rightvsarrow \hspace{-0.02cm} I})^2 + \sin^2 \phi_{t,A}^{S \rightvsarrow \hspace{-0.02cm} I}})$. 
In the same manner, the modification elevation angle $\tilde{\phi}_{t,D}^{I \rightvsarrow \hspace{-0.02cm} G}(\pmb{\psi}_{t,D}^{I \rightvsarrow \hspace{-0.02cm}G}, \eta_{t})$ of IRS-GU link can be obtained as
\begin{eqnarray}\label{app:Omega_G}
&&\hspace{-1.0cm}\tilde{\phi}_{t,D}^{I \rightvsarrow \hspace{-0.02cm} G}\hspace{-0.05cm}(\pmb{\psi}_{t,D}^{I \rightvsarrow \hspace{-0.02cm}G}, \eta_{t}\hspace{-0.05cm}) \hspace{-0.1cm}=\hspace{-0.1cm} \tan^{-1}\hspace{-0.13cm}\left(\hspace{-0.13cm} \frac{\sin(\eta_t-\alpha_{t,D}^{I \rightvsarrow \hspace{-0.02cm} G})}{\sqrt{B_{t,D}^{I \rightvsarrow \hspace{-0.02cm} G}+\cos^2 (\eta_t-\alpha_{t,D}^{I \rightvsarrow \hspace{-0.02cm} G})}}\hspace{-0.13cm}\right)\hspace{-0.12cm},
\end{eqnarray}
where\hspace{-0.02cm} $\alpha_{t,D}^{I \rightvsarrow \hspace{-0.02cm} G}\hspace{-0.12cm}=\hspace{-0.12cm}\tan^{-1}\left( {\tan \phi_{t,D}^{I \rightvsarrow \hspace{-0.02cm} G}}\hspace{-0.04cm}/\hspace{-0.03cm}{\sin \theta_{t,D}^{I \rightvsarrow \hspace{-0.02cm} G}} \right)$ and $B_{t,D}^{I \rightvsarrow \hspace{-0.02cm} G}\hspace{-0.12cm}=\hspace{-0.12cm}(\cos \phi_{t,D}^{I \rightvsarrow \hspace{-0.02cm} G}$ $\cos \theta_{t,D}^{I \rightvsarrow \hspace{-0.02cm} G})^2 /\left({(\cos \phi_{t,D}^{I \rightvsarrow \hspace{-0.02cm} G} \sin \theta_{t,D}^{I \rightvsarrow \hspace{-0.02cm} G})^2 + \sin^2 \phi_{t,D}^{I \rightvsarrow \hspace{-0.02cm} G}}\right)$.
By substituting $\phi_{t,A}^{S \rightvsarrow \hspace{-0.02cm} I}$ and $ \phi_{t,D}^{I \rightvsarrow \hspace{-0.02cm} G}$ in (\ref{eq:arrayresp_I_1}) and (\ref{eq:arrayresp_I_2}) with $\tilde{\phi}_{t,A}^{S \rightvsarrow \hspace{-0.02cm} I}$ and $\tilde{\phi}_{t,D}^{I \rightvsarrow \hspace{-0.02cm} G}$ in (\ref{app:Omega_S}) and (\ref{app:Omega_G}), we derive the array response of tilted IRS as 
\begin{eqnarray}\label{eq:arrayresp_M_1}
&&\hspace{-1.4cm}\tilde{\pmb{a}}_{t,A}^{S \rightvsarrow \hspace{-0.02cm} I}(\pmb{\psi}_{t,A}^{S \rightvsarrow \hspace{-0.02cm}I}, \eta_{t}) = \pmb{v}(d_{x,I} \cos \tilde{\phi}_{t,A}^{S \rightvsarrow \hspace{-0.02cm} I} \cos \tilde{\theta}_{t,A}^{S \rightvsarrow \hspace{-0.02cm} I},  M_{x}) \nonumber\\
&&\hspace{+0.5cm}\otimes\hspace{0.1cm} \pmb{v}(d_{y,I} \cos \tilde{\phi}_{t,A}^{S \rightvsarrow \hspace{-0.02cm} I} \sin\tilde{\theta}_{t,A}^{S \rightvsarrow \hspace{-0.02cm} I},  M_{y}) \in \mathbb{C}^{M\times1},\\
&&\hspace{-1.4cm}\tilde{\pmb{a}}_{t,D}^{I \rightvsarrow \hspace{-0.02cm} G}(\pmb{\psi}_{t,D}^{I \rightvsarrow \hspace{-0.02cm}G}, \eta_{t}) = \pmb{v}(d_{x,I} \cos \tilde{\phi}_{t,D}^{I \rightvsarrow \hspace{-0.02cm} G} \cos \tilde{\theta}_{t,D}^{I \rightvsarrow \hspace{-0.02cm} G},  M_{x})  \nonumber\\
&&\hspace{+0.5cm}\otimes\hspace{0.1cm}\pmb{v}(d_{y,I} \cos \tilde{\phi}_{t,D}^{I \rightvsarrow \hspace{-0.02cm} G} \sin \tilde{\theta}_{t,D}^{I \rightvsarrow \hspace{-0.02cm} G},  M_{y}) \in \mathbb{C}^{M\times1},\label{eq:arrayresp_M_2}
\end{eqnarray}
where $\tilde{\theta}_{t,A}^{S \rightvsarrow \hspace{-0.02cm} I}$$=$$\tan^{-1}\left(\tan{\theta}_{t,A}^{S \rightvsarrow \hspace{-0.02cm} I} \cos \eta_t + \tan{\phi}_{t,A}^{S \rightvsarrow \hspace{-0.02cm} I}\sec{\theta}_{t,A}^{S \rightvsarrow \hspace{-0.02cm} I} \sin \eta_t \right)$ and  $\tilde{\theta}_{t,D}^{I \rightvsarrow \hspace{-0.02cm} G}$$=$$\tan^{-1}\left(\tan{\theta}_{t,D}^{I \rightvsarrow \hspace{-0.02cm} G} \cos \eta_t + \tan{\phi}_{t,D}^{I \rightvsarrow \hspace{-0.02cm} G}\sec{\theta}_{t,D}^{I \rightvsarrow \hspace{-0.02cm} G} \sin \eta_t \right)$.
Consequently, the channel between LEO satellite and IRS tilted by $\eta_t$ can be written as
\begin{eqnarray}\label{eq:channel_IS}
&&\hspace{-1.9cm}\pmb{H}^{S \rightvsarrow \hspace{-0.02cm} I}_{t} (\pmb{\psi}^{S \rightvsarrow \hspace{-0.02cm} I}_t,  \eta_{t})=\beta^{S \rightvsarrow \hspace{-0.02cm} I}_t e^{-jk_w d^{S\text{-} I}_t}
\nonumber\\
&&\hspace{+1.0cm}\times \tilde{\pmb{a}}_{t,A}^{S \rightvsarrow \hspace{-0.02cm} I}(\pmb{\psi}_{t,A}^{S \rightvsarrow \hspace{-0.02cm}I}, \eta_{t})
\left(\pmb{a}_{t,D}^{S \rightvsarrow \hspace{-0.02cm} I}(\pmb{\psi}_{t,D}^{S \rightvsarrow \hspace{-0.02cm}I})\right)^T\hspace{-0.2cm},
\end{eqnarray}
where $\pmb{\psi}^{S \rightvsarrow \hspace{-0.02cm} I}_t \hspace{-0.05cm}=\hspace{-0.05cm} \{\pmb{\psi}_{t,D}^{S \rightvsarrow \hspace{-0.02cm}I}, \pmb{\psi}_{t,A}^{S \rightvsarrow \hspace{-0.02cm}I}\}$, $d^{S\text{-} I}_t$ is the distance between LEO satellite and IRS, and $\beta^{S \rightvsarrow \hspace{-0.02cm} I}_t$ indicates the channel gain of the LEO satellite-IRS link.
Also, the channel between IRS and GU is similarly given as
\begin{eqnarray}\label{eq:channel_GI}
&&\hspace{-1.8cm}\pmb{H}^{I\rightvsarrow \hspace{-0.02cm} G}_{t}(\pmb{\psi}^{I\rightvsarrow \hspace{-0.02cm} G}_t, \eta_{t})= \beta^{I \rightvsarrow \hspace{-0.02cm} G}_t e^{-jk_w d^{I\text{-} G}_t}\nonumber \\
&&\hspace{+0.9cm} \times \pmb{a}_{t,A}^{I \rightvsarrow \hspace{-0.02cm} G}(\pmb{\psi}_{t,A}^{I \rightvsarrow \hspace{-0.02cm}G}) \left(\tilde{\pmb{a}}_{t,D}^{I \rightvsarrow \hspace{-0.02cm} G}(\pmb{\psi}_{t,D}^{I \rightvsarrow \hspace{-0.02cm}G}, \eta_{t}) \right)^T\hspace{-0.2cm},
\end{eqnarray}
where $\pmb{\psi}^{I \rightvsarrow \hspace{-0.02cm} G}_t \hspace{-0.05cm}=\hspace{-0.05cm}\{\pmb{\psi}_{t,D}^{I \rightvsarrow \hspace{-0.02cm}G}, \pmb{\psi}_{t,A}^{I \rightvsarrow \hspace{-0.02cm}G}\}$, $d^{I\text{-} G}$ represents the fixed distance between GU and IRS, and $\beta^{I \rightvsarrow \hspace{-0.02cm} G}_t$ indicates the channel gain of IRS-GU link.
Here, we define the effective channel gain of the LEO satellite-IRS-GU link as
\begin{eqnarray}\label{eq:channel_beta}
&&\hspace{-1.2cm}\beta^{S\rightvsarrow \hspace{-0.02cm} G, \text{eff}}_{t}(\eta_t) \triangleq \beta^{S \rightvsarrow \hspace{-0.02cm} I}_{t} \beta^{I \rightvsarrow \hspace{-0.02cm} G}_{t} = \delta^{S\rightvsarrow \hspace{-0.02cm} G, \text{eff}}_{t} \sqrt{F(\pmb{\psi}_{t,A}^{S \rightvsarrow \hspace{-0.02cm}I}, \pmb{\psi}_{t,D}^{I \rightvsarrow \hspace{-0.02cm}G}, \eta_t)}
\end{eqnarray}
where we have defined $\delta^{S\rightvsarrow \hspace{-0.02cm} G, \text{eff}}_{t}\hspace{-0.05cm} $$=$$ \hspace{-0.05cm}\sqrt{{G_G G_S G_I d_{x\text{,}I} d_{y\text{,}I} \lambda^2}}/$ $\sqrt{{64 \pi^3 (d^{S\text{-} I}_{t} d^{I\text{-} G})^2}}$, $G_G, G_S\hspace{+0.1cm} \text{and}\ G_I$ denote the gains of GU, LEO satellite and IRS,  respectively \cite{Cheng22TC}, and $F(\pmb{\psi}_{t,A}^{S \rightvsarrow \hspace{-0.02cm}I}, \pmb{\psi}_{t,D}^{I \rightvsarrow \hspace{-0.02cm}G}, \eta_t)$ is the power radiation pattern of LEO satellite-IRS-GU link defined as
\begin{eqnarray}\label{eq:channel_power}
&&\hspace{-0.7cm}F(\hspace{-0.02cm}\pmb{\psi}_{t,A}^{S \rightvsarrow \hspace{-0.02cm}I}, \pmb{\psi}_{t,D}^{I \rightvsarrow \hspace{-0.02cm}G}, \eta_t\hspace{-0.02cm})\hspace{-0.1cm}=\hspace{-0.1cm}\sin^{k} \hspace{-0.1cm}\left( \hspace{-0.05cm}\tilde{\phi}_{t,A}^{S \rightvsarrow \hspace{-0.02cm} I}\hspace{-0.02cm}(\hspace{-0.02cm}\pmb{\psi}_{t,A}^{S \rightvsarrow \hspace{-0.02cm}I}, \eta_{t}\hspace{-0.02cm})\hspace{-0.02cm}\hspace{-0.05cm} \right) \hspace{-0.05cm} \sin^{k} \hspace{-0.1cm}\left(\hspace{-0.05cm}\tilde{\phi}_{t,D}^{I \rightvsarrow \hspace{-0.02cm} G}\hspace{-0.02cm}(\hspace{-0.02cm}\pmb{\psi}_{t,D}^{I \rightvsarrow \hspace{-0.02cm}G}, \eta_{t}\hspace{-0.02cm})\hspace{-0.02cm}\hspace{-0.05cm}\right)\hspace{-0.1cm},\nonumber\\
\end{eqnarray}
\vspace{-0.7cm}\\
where $k$ is a directional coefficient of IRS. Similarly, the direct link between GU and LEO satellite can be expressed as
\begin{eqnarray}\label{eq:channel_GS}
&&\hspace{-1.9cm}\pmb{H}^{S\rightvsarrow \hspace{-0.02cm} G}_{t}(\pmb{\psi}^{S \rightvsarrow \hspace{-0.02cm} G}_{t})=\beta^{S \rightvsarrow \hspace{-0.02cm} G}_{t}e^{-jk_w d^{S-G}_{t}} \nonumber \\
&&\hspace{+1.0cm}\times \pmb{a}_{t,A}^{S \rightvsarrow \hspace{-0.02cm} G}(\pmb{\psi}_{t,A}^{S \rightvsarrow \hspace{-0.02cm}G}) 
 \left(\pmb{a}_{t,D}^{S \rightvsarrow \hspace{-0.02cm} G}(\pmb{\psi}_{t,D}^{S \rightvsarrow \hspace{-0.02cm}G})\right)^T\hspace{-0.2cm},
\end{eqnarray}
where $\pmb{\psi}^{S \rightvsarrow \hspace{-0.02cm} G}_{t}\hspace{-0.3cm}=\hspace{-0.3cm}\{\pmb{\psi}_{t,D}^{S \rightvsarrow \hspace{-0.02cm}G}, \pmb{\psi}_{t,A}^{S \rightvsarrow \hspace{-0.02cm}G}
\}$, and we have defined channel amplitude gain from LEO satellite to GU as
$\beta^{S \rightvsarrow \hspace{-0.02cm} G}_{t}\hspace{-0.3cm} =\hspace{-0.3cm} {\sqrt{G_G G_S \lambda^2}}\sin^{k_t/2}\hspace{-0.02cm}(\phi_{t,D}^{S\rightvsarrow \hspace{-0.02cm} G})$ $\sin^{k_r/2}\hspace{-0.02cm}(\phi_{t,A}^{S\rightvsarrow \hspace{-0.02cm} G})/4 \pi d^{S\text{-} G}_{t}$ with $k_t$ and $k_r$ defining as the directional coefficient of LEO satellite and GU, respectively, and $d^{S\text{-} G}_{t}$ as the distance between them.

%%%%%%%%%%%%%%%%%%%%%%%%%%%%%%%%%%%%%%%%%%%%%%%%%%%%%
%% Section III

\section{Optimization of IRS-aided LEO satellite communication systems}\label{sec:solution}
%%%%%%%%%%%%%%%%%%%%%%%%%%%%%%%%%%%%%%%%%%%%%%%%%%%%%
%% Section  III-A
Here, we formulate the problems and develop the solutions of jointly optimizing the mechanical and electrical design of IRS and transceiver beamforming according to the existence of direct link between LEO satellite and GU. We aim to maximize the effective channel gain for two scenarios in Fig. \ref{fig:sys_1}.
In Scenario I without direct link, the overall effective channel matrix of LEO satellite-GU link at time $t$ can be given as
\vspace{-0.1cm}
\begin{eqnarray}\label{eq:channel_1}
&&\hspace{-1.3cm}\pmb{H}^{\text{Sce.I}}_{t}(\pmb{\psi}^{\text{Sce.I}}_{t}, \pmb{\Theta}_{t}, \eta_{t})\hspace{-0.05cm}=\hspace{-0.05cm}\pmb{H}^{I\rightvsarrow \hspace{-0.02cm} G}_{t}(\pmb{\psi}^{I \rightvsarrow \hspace{-0.02cm} G}_{t}, \eta_{t}) \pmb{\Theta}_{t}
\pmb{H}^{S \rightvsarrow \hspace{-0.02cm} I}_{t} (\pmb{\psi}^{S \rightvsarrow \hspace{-0.02cm} I}_{t},  \eta_{t}),
\end{eqnarray}
where $\pmb{\psi}^{\text{Sce.I}}_{t}$$=$$\{\pmb{\psi}^{S \rightvsarrow \hspace{-0.02cm} I}_{t}, \pmb{\psi}^{I \rightvsarrow \hspace{-0.02cm} G}_{t} \}$.
The overall effective channel matrix in Scenario II with direct link is written as
\vspace{-0.1cm}
\begin{eqnarray}\label{eq:channel_2}
&&\hspace{-0.7cm}\pmb{H}^{\text{Sce.II}}_{t} (\pmb{\psi}^{\text{Sce.II}}_{t}, \pmb{\Theta}_{t}, \eta_{t}) \hspace{-0.05cm}=\hspace{-0.05cm} \pmb{H}^{S\rightvsarrow \hspace{-0.02cm}G}_{t}(\pmb{\psi}^{S \rightvsarrow \hspace{-0.02cm} G}_{t})+\pmb{H}^{\text{Sce.I}}_{t} (\pmb{\psi}^{\text{Sce.I}}_{t}, \pmb{\Theta}_{t}, \eta_{t}),\nonumber\\
\end{eqnarray}
\vspace{-0.7cm}\\
where $\pmb{\psi}^{\text{Sce.II}}_{t} = \{\pmb{\psi}^{S \rightvsarrow \hspace{-0.02cm} G}_{t}, \pmb{\psi}^{\text{Sce.I}}_{t} \}$.
Finally, we have the optimization problem as
\vspace{-0.2cm}
\begin{subequations}\label{eq:opt12}
\begin{eqnarray}
&&\hspace{-2.2cm} \max_{\pmb{w}^G_{t}, \pmb{w}^S_{t}, \pmb{\Theta}_{t}, \eta_{t}} \hspace{0cm} \left\lvert \left({\pmb{w}^G_{t}}\right)^{T} \pmb{H}^{\mu}_{t}\left( \pmb{\psi}^{\mu}_{t}, \pmb{\Theta}_{t}, \eta_{t} \right) \pmb{w}^S_{t}\right\rvert ^2 \label{eq:obj12}\\
&&\hspace{-1.1cm} \text{s.t.} \hspace{0.2cm} \lvert \theta_t^{m} \rvert = 1,  \hspace{0.2cm} \forall m \in \{1,...,M\},\label{eq:phase}\\
&&\hspace{-0.5cm} \lvert\lvert\pmb{w}^G_{t} \rvert\rvert ^2 = 1, \hspace{0.2cm} \lvert\lvert\pmb{w}^S_{t} \rvert\rvert ^2 = 1, \label{eq:beamGS}\\
&&\hspace{-0.4cm} \alpha_{t,D}^{I\rightvsarrow \hspace{-0.02cm} G} \le \eta_{t} \le \alpha_{t,A}^{S\rightvsarrow \hspace{-0.02cm} I} \label{eq:eta},
\end{eqnarray}
\end{subequations}
where $\mu \in \{\text{Sce.I}, \text{Sce.II}\}$ is the indicator for differentiating Scenario I and II, and $\pmb{w}^S_{t} \in \mathbb{C}^{N_S \times 1}$ and $\pmb{w}^G_{t} \in \mathbb{C}^{N_G \times 1}$ represent transmit and receive beamforming vectors of LEO satellite and GU, respectively.
In (\ref{eq:opt12}), (\ref{eq:beamGS}) defines the constraints for the normalized beamforming vector, and (\ref{eq:eta}) outlines the constraint for the available range of tilting angle of IRS ensuring visibility of both LEO satellite and GU. Since (\ref{eq:obj12}) is the square of bilinear form, (\ref{eq:opt12}) is non-convex. To address the non-convexity of (\ref{eq:opt12}), we develop the cooperative beamforming design with common phase shift between IRS and GU, and utilize singular value decomposition (SVD) for transmit beamforming design in LEO satellite, whose details are provided in the following.
\vspace{-0.3cm}
\subsection{Optimal design for Scenario I without direct link}\label{subsec:scenario1}
%\vspace{-0.1cm}
\hspace{-0.2cm}In Scenario I, the objective function (\ref{eq:obj12}) can be decomposed as 
\vspace{-0.2cm}
\begin{eqnarray}\label{eq:channel_1_1}
&&\hspace{-0.65cm}
\left(\beta^{S\rightvsarrow \hspace{-0.02cm} G, \text{eff}}_{t}(\eta_t)\right)^{\hspace{-0.1cm}2}
\underbrace{\left\lvert \left(\pmb{w}^G_{t}\right)^{T}\pmb{a}_{t,A}^{I \rightvsarrow \hspace{-0.02cm} G}(\pmb{\psi}_{t,A}^{I \rightvsarrow \hspace{-0.02cm}G}) \right\rvert^2}_{\pmb{h}^G\left( \pmb{w}^G_{t}\right)} \underbrace{\left\lvert  \left(\pmb{a}_{t,D}^{S \rightvsarrow \hspace{-0.02cm} I}\left(\pmb{\psi}_{t,D}^{S \rightvsarrow \hspace{-0.02cm}I}\right)
\right)^T  \pmb{w}^S_{t} \right\rvert^2}_{\pmb{h}^S \left(\pmb{w}^S_{t} \right)}\nonumber\\
&&\hspace{+0.1cm}\times\underbrace{ \left\lvert \varepsilon_G \varepsilon_S \left(\tilde{\pmb{a}}_{t,D}^{I \rightvsarrow \hspace{-0.02cm} G}(\pmb{\psi}_{t,D}^{I \rightvsarrow \hspace{-0.02cm}G}, \eta_{t})
\right)^T \pmb{\Theta}_t  \tilde{\pmb{a}}_{t,A}^{S \rightvsarrow \hspace{-0.02cm} I}(\pmb{\psi}_{t,A}^{S \rightvsarrow \hspace{-0.02cm}I}, \eta_{t}) \right\rvert^2 }_{\pmb{h}^I\left(\pmb{\Theta}_t, \eta_t \right)}. 
\end{eqnarray}
with defining $\pmb{h}^G (\pmb{w}^G_{t})$ on the GU side, $\pmb{h}^S(\pmb{w}^S_{t})$ on the satellite side and $\pmb{h}^I(\pmb{\Theta}_t, \eta_t )$ on the IRS side, where $\varepsilon_G$$=$$e^{-jk_wd^{I\text{-} G}}$ and $\varepsilon_S$$=$$e^{-jk_wd^{S\text{-} I}_{t}}$. 
For the optimal tilt angle $\eta_{t,\text{opt}}$ to satisfy the maximum of squared effective channel gain $(\hspace{-0.02cm}\beta^{S\rightvsarrow \hspace{-0.02cm} G, \text{eff}}_{t}\hspace{-0.02cm}(\eta_t)\hspace{-0.02cm})^2$, we can obtain the stationary point of $F(\pmb{\psi}_{t,A}^{S \rightvsarrow \hspace{-0.02cm}I}, \pmb{\psi}_{t,D}^{I \rightvsarrow \hspace{-0.02cm}G}, \eta_t)$ as the optimal solution of an unconstrained convex optimization problem \cite{Boyd04Convex} given by 
\vspace{-0.1cm}
\begin{eqnarray}
&&\hspace{-1.6cm} \eta_{t, \text{opt}}= \arg \max_{\eta_t} \hspace{0.1cm}F(\pmb{\psi}_{t,A}^{S \rightvsarrow \hspace{-0.02cm}I}, \pmb{\psi}_{t,D}^{I \rightvsarrow \hspace{-0.02cm}G}, \eta_t) = \frac{\alpha_{t,A}^{S \rightvsarrow \hspace{-0.02cm} I}+\alpha_{t,D}^{I \rightvsarrow \hspace{-0.02cm} G}}{2}.
\label{eq:opt_tilt}
\end{eqnarray}
With the common phase-shift $\rho_t$ between IRS and GU, the optimal passive beamforming vector for IRS to maximize $\pmb{h}^I(\pmb{\Theta}_t, \eta_t)$ can be calculated by aligning the phase through the Hadamard product as
\vspace{-0.2cm}
\begin{eqnarray}\label{eq:passivebeam}
&&\hspace{-1.1cm} \pmb{\Theta}_{t, \text{opt}} = e^{j \rho_t} \left(\varepsilon_G \varepsilon_S  \tilde{\pmb{a}}_{t,D}^{I \rightvsarrow \hspace{-0.02cm} G}(\pmb{\psi}_{t,D}^{I \rightvsarrow \hspace{-0.02cm}G}, \eta_{t})  \odot \tilde{\pmb{a}}_{t,A}^{S \rightvsarrow \hspace{-0.02cm} I}(\pmb{\psi}_{t,A}^{S \rightvsarrow \hspace{-0.02cm}I}, \eta_{t})  \right)^{*} \hspace{-0.05cm},
\end{eqnarray}
where $\rho_{t}$$=$$\rho^{\text{Sce.I}}_{t,\text{opt}}$$=$$0$ as the optimal common phase-shift for Scenario I. Then, in order to maximize $\pmb{h}^G(\pmb{w}^G_{t})$ and $\pmb{h}^S(\pmb{w}^S_{t})$, we propose the use of maximum-ratio transmission and combination (MRT/MRC) as beamforming solution \cite{Bjornson13Optimal}, i.e.,
\begin{eqnarray}\label{eq:MRT/MRC}
&&\hspace{-1.3cm} \pmb{w}_{t, \text{opt}}^{G,\text{Sce.I}} \hspace{-0.1cm}=\hspace{-0.1cm} \frac{\left(\pmb{a}_{t,A}^{I \rightvsarrow \hspace{-0.02cm} G}(\pmb{\psi}_{t,A}^{I \rightvsarrow \hspace{-0.02cm}G})\right)^*}{\lvert\lvert \pmb{a}_{t,A}^{I \rightvsarrow \hspace{-0.02cm} G}(\pmb{\psi}_{t,A}^{I \rightvsarrow \hspace{-0.02cm}G}) \rvert\rvert}\hspace{+0.2cm} \text{and} \hspace{+0.2cm}
\pmb{w}_{t,\text{opt}}^{S,\text{Sce.I}}\hspace{-0.1cm}=\hspace{-0.1cm}\frac{\left(\pmb{a}_{t,D}^{S \rightvsarrow \hspace{-0.02cm} I}(\pmb{\psi}_{t,D}^{S \rightvsarrow \hspace{-0.02cm}I})\right)^*}{\lvert\lvert \pmb{a}_{t,D}^{S \rightvsarrow \hspace{-0.02cm} I}(\pmb{\psi}_{t,D}^{S \rightvsarrow \hspace{-0.02cm}I}) \rvert\rvert}.
\end{eqnarray}
\vspace{-0.5cm}
\subsection{Optimal solution for Scenario II}\label{subsec:scenario2}
In Scenario II with a direct link, the objective function (\ref{eq:obj12}) upon the optimal passive beamforming $\pmb{\Theta}_{t, \text{opt}}$ in (\ref{eq:passivebeam}) can be decomposed as 
\vspace{-0.3cm}
\begin{eqnarray}
&&\hspace{-0.7cm}\underbrace{ \Big\lvert \left( \pmb{w}^G_{t}\right)^T 
\overbrace{
\left[\beta^{S \rightvsarrow \hspace{-0.02cm} G}_t \varepsilon_{SG},  \beta^{S\rightvsarrow \hspace{-0.02cm} G, \text{eff}}_{t}(\eta_t) M e^{j\rho_t}\right] 
\begin{bmatrix}
\pmb{a}_{t,A}^{S \rightvsarrow \hspace{-0.02cm} G}(\pmb{\psi}_{t,A}^{S \rightvsarrow \hspace{-0.02cm}G}) \\
\pmb{a}_{t,A}^{I \rightvsarrow \hspace{-0.02cm} G}(\pmb{\psi}_{t,A}^{I \rightvsarrow \hspace{-0.02cm}G})
\end{bmatrix}}^{\pmb{H}^G}
\Big\rvert^2}_{\pmb{g}^G(\pmb{w}_t^G,\rho_t, \eta_t)}\nonumber
\end{eqnarray}
\vspace{-0.78cm}
\begin{eqnarray}
&&\hspace{-5.0cm}\times\underbrace{ \Big\lvert
\overbrace{
\begin{bmatrix}
\pmb{a}_{t,D}^{S \rightvsarrow \hspace{-0.02cm} G}(\pmb{\psi}_{t,D}^{S \rightvsarrow \hspace{-0.02cm}G}) \\
\pmb{a}_{t,D}^{S \rightvsarrow \hspace{-0.02cm} I}(\pmb{\psi}_{t,D}^{S \rightvsarrow \hspace{-0.02cm}I})
\end{bmatrix}}^{\pmb{H}^S}
 \pmb{w}_t^S
\Big\rvert^2}_{\pmb{g}^S(\pmb{w}_t^S)},
\end{eqnarray}
where we have defined $\pmb{g}^G(\pmb{w}_t^G, \rho_t, \eta_t)$ with the channel matrix $\pmb{H}^G$ on the GU side and $\varepsilon_{SG}$$=$$e^{-jk_wd^{S\text{-} G}_{t}}$, and $\pmb{h}_2(\pmb{w}_t^S)$ with the channel matrix $\pmb{H}^S$ on the satellite side. Here, the optimal tilt angle $\eta_{t,\text{opt}}$ of IRS to maximize $\beta^{S\rightvsarrow \hspace{-0.02cm} G, \text{eff}}_{t}(\eta_t)$ can be induced as the same solution of (\ref{eq:opt_tilt}). With the optimal beamforming of GU side as $\pmb{w}_{t, \text{opt}}^{G,\text{Sce.II}}$$=$$(\pmb{H}^G)^* / \vert\vert \pmb{H}^G\vert\vert^2$, $\pmb{g}^G(\pmb{w}_t^G, \rho_t, \eta_t)$ can be reformulated as
\begin{eqnarray}\label{eq:channel_g_1}
&&\hspace{-0.7cm} \pmb{g}^G(\pmb{w}_t^G, \rho_t, \eta_t) = \lvert \beta^{S \rightvsarrow \hspace{-0.02cm} G}_{t} \rvert^2 N_G + \left(\beta^{S\rightvsarrow \hspace{-0.02cm} G, \text{eff}}_{t}(\eta_t)\right)^2 N_G M^2 \nonumber \\
&&\hspace{-0.4cm}  + 2 M \beta^{S \rightvsarrow \hspace{-0.02cm} G}_{t} \left(\beta^{S\rightvsarrow \hspace{-0.02cm} G, \text{eff}}_{t}(\eta_t)\right)^2\nonumber\\
&&\hspace{-0.4cm} + \text{Re} \hspace{-0.1cm}\left[e^{j\rho_t}\varepsilon^{*}_{SG} \left(\tilde{\pmb{a}}_{t,D}^{I \rightvsarrow \hspace{-0.02cm} G}(\pmb{\psi}_{t,D}^{I \rightvsarrow \hspace{-0.02cm}G}, \eta_{t}) \right)^{H} \hspace{-0.1cm} \pmb{a}_{t,A}^{S \rightvsarrow \hspace{-0.02cm} G}(\pmb{\psi}_{t,A}^{S \rightvsarrow \hspace{-0.02cm}G})  \right],
\end{eqnarray}
where $\text{Re} [\cdot]$ denotes the function that extracts the real part of element.
At this point, the optimal common phase-shift for the case of Scenario II can be determined as $\rho^{\text{Sce.II}}_{t,\text{opt}}$$=$$\angle(\varepsilon_{SG})-\angle ( (\pmb{a}_{t,A}^{S \rightvsarrow \hspace{-0.02cm} G}(\pmb{\psi}_{t,A}^{S \rightvsarrow \hspace{-0.02cm}G}))^H  \pmb{a}_{t,A}^{I \rightvsarrow \hspace{-0.02cm} G}(\pmb{\psi}_{t,A}^{I \rightvsarrow \hspace{-0.02cm}G}) )$, where $\angle(\cdot)$ denotes the phase of the element. To maximize $\pmb{g}^S(\pmb{w}_t^S)$, the optimal transmit beamforming in satellite side $\pmb{w}_{t, \text{opt}}^{S,\text{Sce.II}}$ can be obtained by using the SVD of $\pmb{H}^S$, i.e., $\pmb{H}^S$$=$$\pmb{U}_S\pmb{\Sigma}_S (\pmb{V}_S)^H$. By reformulating $\pmb{g}^S(\pmb{w}_t^S)$ as $\vert \pmb{\Sigma}_S(\pmb{V}_S)^H \pmb{w}_t^S \vert^2$, the optimal transmit beamforming can be induced by aligning $\pmb{w}_t^S$ with the direction of the largest singular value, corresponding to the first column of $\pmb{V}_S$.
\begin{lemma}
When the angular deviation between the LEO satellite-GU link and the LEO satellite-IRS link on the satellite side is minimal, leading to $\pmb{a}_{t,D}^{S \rightvsarrow \hspace{-0.02cm} G}(\pmb{\psi}_{t,D}^{S \rightvsarrow \hspace{-0.02cm}G}) \approx \pmb{a}_{t,D}^{S \rightvsarrow \hspace{-0.02cm} I}(\pmb{\psi}_{t,D}^{S \rightvsarrow \hspace{-0.02cm}I})$, the same transmit beamforming solution of LEO satellite in (\ref{eq:MRT/MRC}) also can be applied for Scenario II.
\end{lemma}

    \vspace{-0.2cm}
\section{Numerical Results}\label{sec:num}
 
In this section, we evaluate the performances of the proposed scheme for the IRS-aided LEO satellite MIMO systems. Following the real orbit data of the Orbcomm LEO satellite \cite{Kassas23PLANS}, we consider the LEO satellite orbiting the Earth, which has a radius of $R_E$$=$$6371$ km, at a fixed radius of $R_S$$=$$740$ km and a constant speed of $V_S$$=$$7.5$ km/s. For representing dynamic positioning of LEO satellite and position of ground nodes, we geodetic coordinates for all nodes, where the GU is positioned at (51.509°N, 0.009°W, $R_E $$+$$h_G$) heading north along Earth's horizontal tangent on its longitude line, and the IRS is located at (51.512°N, 0°W, $R_E$$+$$h_I$) heading south following the Earth's horizontal tangent on its longitude line.
Unless stated otherwise, by referring \cite{Kassas23PLANS, Matthiesen21WCL, Zheng22JSAC, Tian22arXiv, Zheng22GCWkshps, Cao23PIMRC}, we set the parameters as in Table \ref{sim:parameter}.

For reference, we compare the proposed scheme with several benchmark schemes:
(i) \textit{Optimal 2D $\eta_{t}$ with elevation angles} (Opt. 2D $\eta_{t}$), where the tilt angle $\eta_{t}$ is optimally adjusted in a 2D manner, simplified by considering only the elevation angles as $\eta_{t}$$=$$(\phi_{t,A}^{S\rightvsarrow \hspace{-0.02cm} I}$$-$$\phi_{t,D}^{I\rightvsarrow \hspace{-0.02cm} G})/2$ \cite{Cheng22TWC}.
(ii) \textit{Fixed IRS with $\eta_{t}$$=$$0$} (Fix. $\eta_{t}$$=$$0$), where the IRS's tilt angle $\eta_{t}$ is fixed at 0, which indicates the non-tilted IRS.
(iii) \textit{Isotropic beamforming} (Isotropic beam.), where the IRS is adjusted with optimal electrical adjustment, alongside isotropic transceiver beamforming for GU and LEO satellite.
(iv) \textit{No IRS deployment} (No IRS), where no IRS is deployed in Scenario II, and only the direct link between LEO satellite and GU is considered, with optimal beamforming applied.

\begin{figure}[t!]
\centering
\includegraphics[width=6.5cm]{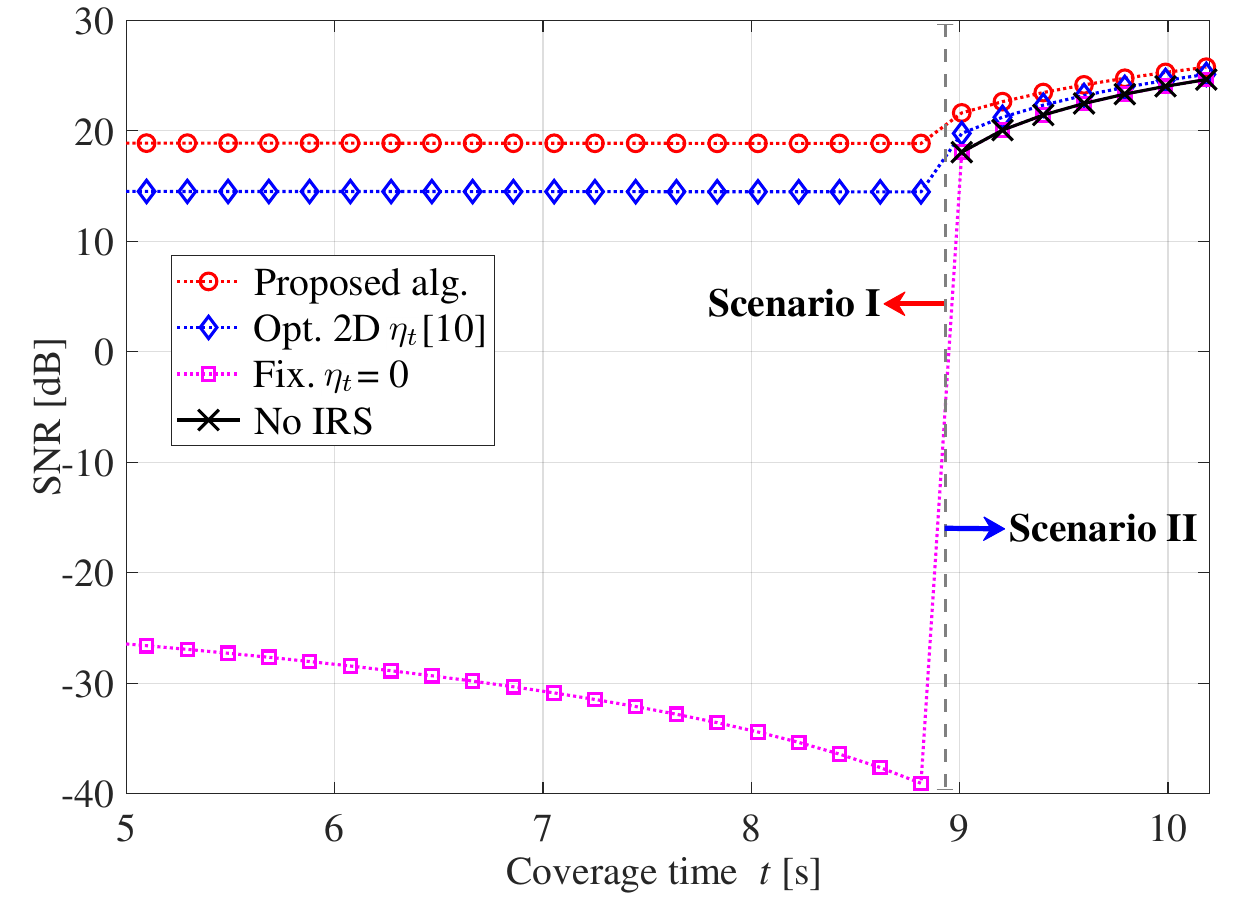}
\caption{Received SNR versus the coverage time $t$ with Scenario I and II.}
\label{fig:sim_1}
    
\end{figure}
\begin{table}[t!]
    \renewcommand\arraystretch{1.5}
    \caption{\centering Simulation parameters} \label{sim:parameter} %title of the table
    \resizebox{\columnwidth}{!}{%
    \begin{tabular}{ c c c c c c } % creating eight columns
    \hline\hline
    Parameter & Value & Parameter & Value & Parameter & Value\\
    \hline 
    $h_I, h_G$ & 150 $\text{m}$, 30 $\text{m}$ & $l_H, l_V$ & 333 m, 623 m & $d_{x,i}, d_{y,i}, i \in \{G,S\}$ & 0.25 m  \\
    $N_{x,i}, N_{y,i},  i \in \{G,S\}$ & 15 & $d_{x,I}, d_{y,I}$ & 0.25 m & $M_x, M_y$ & $20$  \\
    $k_t, k_r, k$ & 1, 1, 2 & $G_G, G_S, G_I$ & 4 dB, 4 dB, 6 dB \cite{Tian22arXiv} & $\lambda $ & 2 m \cite{Kassas23PLANS}\\
    $P_t$ & 15 dBW & $N_0$ & -120 dBW & - & - \\
    \hline\hline
    \label{parameter_table} \end{tabular}}
    \vspace{-0.7cm}
 \end{table}

In Fig. \ref{fig:sim_1}, we illustrate the received SNR of GU versus the coverage time $t$ for both Scenario I and II, where the LEO satellite travels from coordinate (51.49°N, 0.5°W, $R_E $$+$$R_S$) to (51.512°N, 0.5°N, $R_E $$+$$ R_S$).
Due to the severe performance degradation without beamforming design, we do not include the benchmark schemes with isotropic beamforming.
The proposed algorithm can obtain an 30.2\% improvement of 4.38 dB and 9.4\% improvement of 1.86 dB compared to existing 2D tilting design \cite{Cheng22TWC} in Scenario I and II, respectively.
Since LEO satellite communications in Scenario I rely entirely on the virtual link generated by IRS, the proposed algorithm provides higher gain in Scenario I compared to Scenario II with a direct link.
This can be attributed to the angular adjustment on both elevation and azimuth angles, which can be emphasized when the transceiver and IRS are not on the same plane.
Without IRS tilting of $\eta_{t}$$=$$0$, the receive SNR deteriorates significantly as the LEO satellite approaches the intersection of Scenario I and II, 
since the absence of angular adjustment causes $\beta^{S\rightvsarrow \hspace{-0.02cm} G, \text{eff}}_{t}$ to drop with the non-modified elevation angle ${\phi}_{t,A}^{S \rightvsarrow \hspace{-0.02cm} I}$.

\begin{figure}[t!]
\centering
\includegraphics[width=6.5cm]{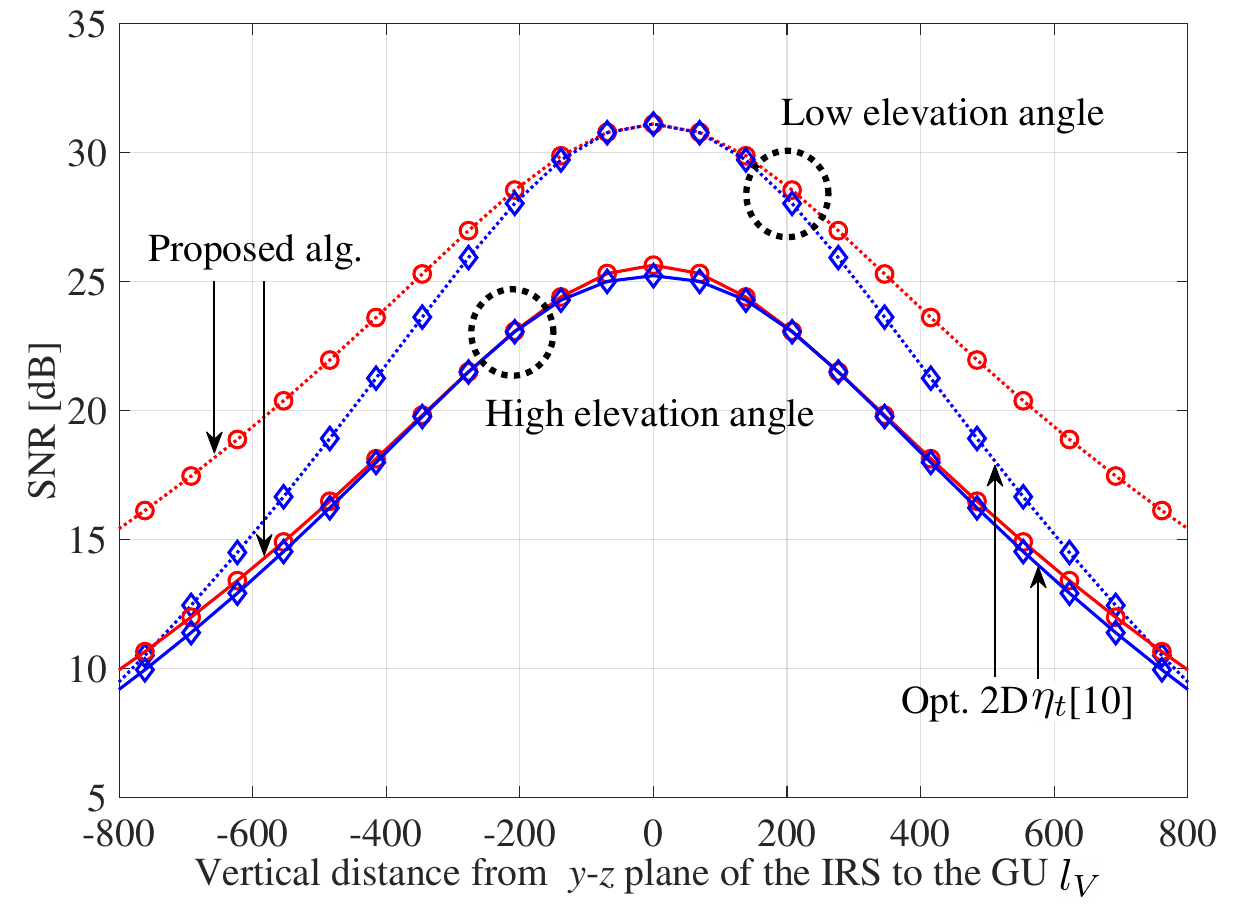}
\caption{Received SNR versus the vertical distance from $y$-$z$ plane of the IRS to the GU $l_V$.}
\label{fig:sim_2}
    \vspace{-0.4cm}
\end{figure}

Fig.\ref{fig:sim_2} illustrates the receive SNR versus the vertical distance $l_V$ (m) from the $y$-$z$ plane of the IRS to the GU of Fig. \ref{fig:sys_1} in Scenario I. To highlight the effects of LEO satellite positions, we classify the position of LEO satellite by $\phi_{t,A}^{S \rightvsarrow \hspace{-0.02cm} I}$: one associated with a low $\phi_{t,A}^{S \rightvsarrow \hspace{-0.02cm} I}$ (Low elevation angle) and the other with a high $\phi_{t,A}^{S \rightvsarrow \hspace{-0.02cm} I}$ (High elevation angle).
For the case of low elevation angle $\pmb{p}_{S,1}\hspace{-0.05cm}=\hspace{-0.05cm}(51.5056\text{°N}, 0.2\text{°W}, R_E \hspace{-0.05cm}+ \hspace{-0.05cm}R_S)$, ($\phi_{t,A}^{S\rightvsarrow \hspace{-0.02cm} I}, \theta_{t,A}^{S\rightvsarrow \hspace{-0.02cm} I}) \hspace{-0.05cm}=\hspace{-0.05cm} (0.06\text{°}, 91.2\text{°})$ is considered, while for the case with high elevation angle $\pmb{p}_{S,2}\hspace{-0.05cm}=\hspace{-0.05cm}(46.30\text{°N}, 15.03\text{°W},  R_E \hspace{-0.03cm}+ \hspace{-0.03cm} R_S)$, ($\phi_{t,A}^{S\rightvsarrow \hspace{-0.02cm} I}, \theta_{t,A}^{S\rightvsarrow \hspace{-0.02cm} I}) \hspace{-0.05cm}=\hspace{-0.05cm}  (20\text{°}, 154\text{°})$ is considered.
In both cases, the highest SNR can be obtained at $l_V$$=$$0$ when the GU is located on the $y$-$z$ plane of the IRS, which is consistent with the results in  \cite{Cheng22TWC}. 
In the case of low elevation angle, $\tilde{\phi}_{t,A}^{S \rightvsarrow \hspace{-0.02cm} I}$ has minimal impact on $\beta^{S\rightvsarrow \hspace{-0.02cm} G, \text{eff}}_{t}$, while the increasing impact of $\tilde{\phi}_{t,D}^{I \rightvsarrow \hspace{-0.02cm} G}$ highlights the difference between $\tilde{\phi}_{t,D}^{I \rightvsarrow \hspace{-0.02cm} G}$ and ${\phi}_{t,D}^{I \rightvsarrow \hspace{-0.02cm} G}$ as $\lvert l_V\rvert$ increases, which emphasizes the difference between optimal tilting and 2D tilting.
In contrast, in the case of high elevation angle, $\tilde{\phi}_{t,A}^{S \rightvsarrow \hspace{-0.02cm} I}$ impacts $\beta^{S\rightvsarrow \hspace{-0.02cm} G, \text{eff}}_{t}$, leading to a performance gap at $l_V$$=$$0$ due to the difference between the $\tilde{\phi}_{t,A}^{S \rightvsarrow \hspace{-0.02cm} I}$ and ${\phi}_{t,A}^{S \rightvsarrow \hspace{-0.02cm} I}$, while as $\lvert l_V\rvert$ increases, the performance gap is caused by the difference between $\tilde{\phi}_{t,D}^{I \rightvsarrow \hspace{-0.02cm} G}$ and ${\phi}_{t,D}^{I \rightvsarrow \hspace{-0.02cm} G}$.

\begin{figure}[t]
\centering
\includegraphics[width=6.5cm]{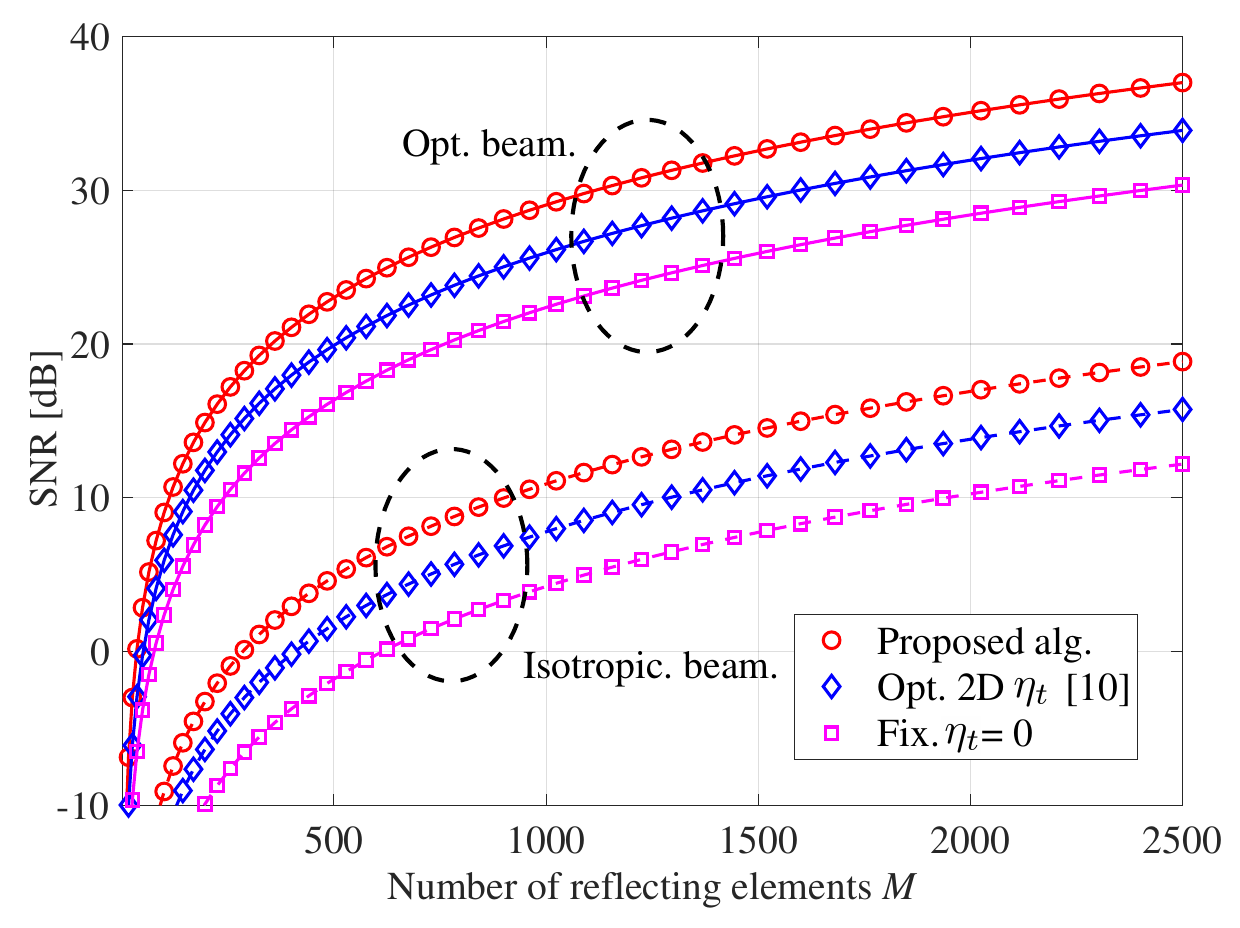}
\caption{Received SNR versus the number of reflecting elements $M$ with high elevation angle.}
\label{fig:sim_3}
    \vspace{-0.4cm}
\end{figure}

Fig. \ref{fig:sim_3} shows the SNR versus the number of reflecting elements $M$ in Scenario I with high elevation angle case.
The proposed algorithm can obtain a performance improvement of 3.11 dB over the 2D tilting optimization and 6.67 dB over without tilting of IRS, regardless of the increasing number of reflecting elements. This consistent improvement can be attributed to the power-scaling law of IRS elements, which implies that the SNR increases proportionally with $M^2$ when employing the optimal IRS design \cite{Cheng22TC}. 
Notably, the optimization of active beamforming for GU and LEO satellites is essential with 18.2 dB improvement over isotropic transceiver beamforming, which becomes enhanced by a factor of $N_G$$\times$$N_S$ with optimal beamforming.

    \vspace{-0.2cm}

\section{Conclusions}\label{sec:con}
In this correspondence, we propose a novel LEO satellite MIMO communication architecture with a aid of a mechanically-tilted IRS. For the general deployment of IRS, we develop the 3D MIMO channel model and consider two possible scenarios according to the presence of the direct link between the GU and the LEO satellite. With the aim of maximizing the overall effective channel gain in LEO satellite communications, we propose the optimal tilting angle and phase shift of the IRS along with the transceiver beamforming for each scenario. 
Via numerical results, the performance enhancement of the proposed algorithm is verified to be pronounced with practical transceiver deployment using real orbit data from the Orbcomm LEO satellite. As future work, the multiple LEO satellites and GUs can be studied with the mechanically-tilted IRS.

    \vspace{-0.2cm}
% \vfill

\end{document}